\documentstyle[subeqn]{article}
\begin{document}
\baselineskip = 18 pt

\begin{titlepage}
\begin{center}
\large {SOLITON SOLUTIONS, LIOUVILLE INTEGRABILITY AND}
\large {GAUGE EQUIVALENCE OF SASA SATSUMA EQUATION}
\end{center}
\vspace{2 cm}
\begin{center}
{\bf Sasanka Ghosh}\footnote{sasanka@iitg.ernet.in}\\
{\it Physics Department, Indian Institute of Technology, Guwahati}\\
{\it Panbazar, Guwahati - 781 001, India}\\
{\bf Anjan Kundu}\footnote{anjan@tnp.saha.ernet.in}\\
{\it TNP Division, Saha Institute of Nuclear Physics}\\
{\it 1/AF Bidhan Nagar, Calcutta - 700 069, India}\\
{\bf Sudipta Nandy}\footnote{sudipta@iitg.ernet.in}\\
{\it Physics Department, Indian Institute of Technology, Guwahati}\\
{\it Panbazar, Guwahati - 781 001, India}\\
\end{center}
\vspace{3 cm}
\begin{abstract}
Exact integrability of the Sasa Satsuma equation (SSE) in the Liouville sense is 
established by showing the existence of an infinite set of conservation 
laws. The explicit form of the conserved quantities in terms of the fields
are obtained by solving the Riccati equation for the associated $3\times 3$
Lax operator. The soliton solutions, in particular, one and two soliton 
solutions, are constructed by the Hirota's bilinear method. The one soliton 
solution is also compared with that found through the inverse scattering 
method. The gauge equivalence of the SSE with a generalized Landau Lifshtz
equation is established with the explicit construction of the new equivalent 
Lax pair.
\end{abstract}
\end{titlepage}
\newpage

\baselineskip = 24 pt

{\bf I. Introduction}\\
Nonlinear Schr\"odinger equation and its various generalized versions (higher 
order nonlinear Schr\"odinger equation) is well known in describing various 
physical phenomena \cite{I}. A common property in all these physical systems 
is the appearance of solitons, as a result of a balance between the 
nonlinear and dispersive terms of the wave equations. With the advancement of 
experimental accuracy, solitons having more complicated dynamics can also be 
detected and observed now \cite{II}. 
The Sasa Satsuma equation \cite{III}
\begin{equation}
 iQ_{T} +\frac{1}{2}Q_{XX}+\vert Q\vert^{2}Q + \frac{i}{6\epsilon}(Q_{XXX} 
+ 6\vert Q\vert^{2}Q_{X} + 3\vert Q\vert_{X}^2 Q ) = 0
\label{II4}
\end{equation}
describing the evolution of a complex scalar field, is an example of such 
a system, whose soliton solutions have been obtained through inverse scattering
method (ISM) in  \cite{III}.

A limited class of soliton bearing equations exhibits further interesting 
properties and belongs to the exclusive club of integrable systems. The 
most prominent definition of integrability is the integrability in the Liouville
sense, i.e., the existence of a set of infinite numbers of conserved quantities in 
involution \cite{IV}, which can be considered as the action variables. 
This criterion of 
integrability is extendable also to the quantum case. The Lax pair associated 
with the model is usually a sign of such integrability, while the Painlev\,{e} 
singularity analysis \cite{V} is supposed to be a direct test of integrability 
for the given equation.

It should be mentioned that the Lax pair for the SSE as well as its one soliton
through ISM were found in \cite{III}, while the Painleve analysis for the equation 
was carried out in \cite{VI}. However, extraction of the higher conserved 
quantities for the Sasa Satsuma system and thus establishing the integrability 
of the whole hierarchy in the Liouville sense remained unexplored. A possible 
reason of this may be the difficulty involved due to the unusual $3\times 3$ 
matrix form of the Lax operator, associated with the SSE. 

Our objective is, therefore, to find the Riccati equaton for the $3\times 3$ Lax 
operator, associated with the SSE and consequently, to obtain the whole 
hierarchy of conserved charges in a systematic way. This constructiom will be 
somewhat involved due to the extended form of the Lax operator. 

For investigating SSE from a different view point, we further find the explicit 
soliton solutions of the equation through Hirota's bilinear method. This is a 
direct and much more effective method compared to ISM for obtaining the soliton 
solutions, since it does not require the knowledge of the Lax pair. Moreover, 
the construction of the  $\tau $ function becomes straightforward in this method. 

One should recall in this context another interesting fact about the NLS equation 
that it is gauge related to the well known Landau Lifshitz equation (LLE) 
\cite{VII}. This equivalence can  also be established  through the space curve 
method \cite{VIII}. It is, therefore, natural to ask what is the gauge equivalent 
equation to the SSE. Though such an equivalent system has been discovered by the
space curve method in \cite{VI}, we complete the investigation for SSE by showing 
the  equivalence of it through an explicit gauge tranformation, which not only 
reproduces the generalized LLE (GLLE), but also constructs the associated Lax 
pair for the GLLE. 

The organization of this paper is as follows. In section II we study the soliton
solutions of SSE by Hirota's bilinear method.  We compute explicitly the one 
and two soliton solutions and compare our result of one soliton solution with
the known one \cite{III}. In section III, we construct the 
related Riccati equation using the $3\times 3$ Lax operator of the SSE and 
subsequently find the infinite number of conserved quantities through the 
recursion relation. The time invariance of the conserved quantities is checked 
directly by using the evolution equation. This proves the Liouville 
integrability of the SSE. The 
section IV, provides the gauge equivalent generalized LLE and gives the 
associated new Lax operators in the explicit form. Section V is the concluding 
one.
\vspace {.5 cm}

{\bf II. Soliton Solutions Through Hirota's Method}

Let us begin with the SSE (\ref{II4}), which through a change of variable and a 
Galelian transformation:
\begin{eqnarray}
 Q(X,T) &=& u(x,t)\exp{\{i\epsilon(x+\frac{\epsilon t}{6})\}}\nonumber\\
 T &=& t\\
 X &=& x+\frac{\epsilon}{2}t\nonumber
\end{eqnarray}
may be simplified to the form
\begin{equation}
u_{t}+\frac{1}{6\epsilon}(u_{xxx}+ 6\mid u\mid ^{2}u_{x}
+ 3(\mid u\mid ^{2})_{x}u) = 0
\label{II1}
\end{equation}
This is an example of a complex modified KdV type equation and goes to mKdV for
the real valued field. The associated spectral 
problem can be studied through the pair of linear equations
\begin{subequations}
\begin{eqnarray}
 \Psi_{x} & = & {\bf U}(x,t,\lambda )\Psi\\
 \Psi_{t} & = & {\bf V}(x,t,\lambda )\Psi
\end{eqnarray} 
\label{II2}
\end{subequations}
where ${\bf U}(x,t,\lambda )$ and ${\bf V}(x,t,\lambda)$ are $3\times 3$ 
matrices and 
$\lambda $ is the spectral parameter. The explicit form of ${\bf U}$ 
and ${\bf V}$ may be given using the result of \cite{III} as
\begin{subequations}
\begin{eqnarray}
 {\bf U} &=& -i\lambda {\bf \Sigma}+ {\bf A}\\ 
 {\bf V} &=& -i4\epsilon \lambda^3 {\bf \Sigma} + 4\epsilon(\lambda^2 - \mid u\mid^2){\bf A}\nonumber \\
& &-i2\epsilon\lambda {\bf \Sigma}({\bf A}^2 -{\bf A}_x) - \epsilon{\bf A}_{xx}
+\epsilon({\bf A}_x{\bf A} - {\bf A}{\bf A}_x) 
\label{II3}
\end{eqnarray}  
\end{subequations}
with 
\[ {\bf \Sigma} = 
\left( \begin{array}{ccc}
            1 & 0 & 0 \\
            0 & 1 & 0 \\
            0 & 0 &-1
\end{array} \right)
\]

\[{\bf A} = \left( \begin{array}{ccc}
           0 & 0 & u \\
           0 & 0 & u^{*} \\
          -u^{*} & -u & 0
 \end{array}\right)\]        
Compatibility of (\ref{II2}a) and (\ref{II2}b) leads to SSE (\ref{II1}), which 
can be shown easily by using the following properties of ${\bf \Sigma }$ and 
$\bf A$ matrices
\begin{eqnarray*}
\{{\bf \Sigma} , {\bf A\}} = 0\\  
{\bf \Sigma} ^2 = 1\\ 
{\bf A}^3 + 2\vert u\vert ^2 {\bf A} = 0 
\end{eqnarray*}
Note that SSE in the form (\ref{II1}), though suitable for studying inverse 
scattering technique, is not convenient for casting it into  Hirota's 
bilinear form. On the other hand, the higher order nonlinear Schrodinger 
equation form (\ref{II4}) for SSE is more suitable for this purpose. Now,
in order to write (\ref{II4}) in the bilinear form, we make the transformation
\begin{equation}
Q(X,T) = G(X,T)/F(X,T)
\label{II5}
\end{equation} 
where, $G$ is complex and $F$ is real. Consequently, in these new variables, 
we have the following set of equations
\begin{subequations}
\begin{eqnarray} 
(iD_{T} +\frac{1}{2}D^2_{X} + \frac{i}{6\epsilon}D^3_X)G.F &=& 0\\ 
D^2_XF.F &=& 4G^{*}.G\\
(1 - \frac{2i}{\epsilon}D_{X})G^{*}.G &=& 0
\end{eqnarray}
\label{II6}
\end{subequations}
which follow from (\ref{II4}). $D_T$, $D_X$, $D_{XX}$ etc. in (\ref{II6})
are Hirota derivatives \cite{IX}. 
(\ref{II6}) belongs to   
a {\it new} class of bilinear equations, whose general form would be of the type
\begin{subequations}
\begin{eqnarray}
{\cal B}(D_{X},D_{T},\cdots)G.F &=& 0\\
{\cal A}(D_{X},D_{T},\cdots)F.F &=& {\cal C}(D_{X},D_{T},\cdots)G^{*}.G\\
{\cal E}(1-D_{X},D_{T},\cdots)G^{*}.G &=& 0
\end{eqnarray}
\end{subequations}
The additional bilinear equation (\ref{II6}c) involving $G^*G$ imposes one further 
condition on the complex parameter $P$ (shown below), which is absent in other 
examples of higher order nonlinear Schr\"odinger equations \cite{X}. 

For obtaining one soliton solution of SSE (\ref{II4}), we choose $G$ and $F$ 
in the following form
\begin{subequations}
\begin{eqnarray}
G = L\exp(\eta )\\
F = 1 + K\exp(\eta +\eta^{*})
\end{eqnarray}
\label{II7}
\end{subequations}
where, $L$ is a complex c-number parameter and 
\begin{equation}
\eta = PX + \Omega T + .......
\label{II8} 
\end{equation}
with $P$, $\Omega $ are in general complex parameters. Substituting the 
expressions (\ref{II7})
for $G$ and $F$ in (\ref{II6}), we see that $G$ and $F$ are the solutions of (\ref{II6}) 
provided the following relations hold
\begin{subequations}
\begin{equation}
i\Omega + \frac{1}{2} P^{2} + i\frac{P^3}{6\epsilon} = 0
\label{II8a}
\end{equation} 
\begin{equation}
 K = \frac{L L^{*}}{2\mu ^{2}} 
\label{II8b}
\end{equation}
\begin{equation} 
P - P^{*} = 2i\epsilon
\label{II8c}
\end{equation}
\end{subequations}
Where the complex parameter, $P$ is of the form
\begin{subequations}
\begin{equation}
P = \mu + i\epsilon 
\label{II9a}
\end{equation}
(\ref{II8a}) is nothing but the dispersion relation and (\ref{II8b}) determines $K$. 
In the above solution, so far, $L$ is an arbitrary complex parameter.
We will see shortly that in order
to compare our result with the one obtained through the ISM 
\cite{III}, the parameter $L$ is to be chosen in a specific form. It follows 
from the dispersion relation (\ref{II8a}) and the expression of $P$ (\ref{II9a}) 
that $\Omega $ should be of the form 
\begin{equation}
\Omega = -\mu (\frac{\mu^2}{6\epsilon } + \frac{\epsilon}{2}) 
-i\frac{\epsilon ^2}{3}
\label{II9b}
\end{equation}
\end{subequations}
Substituting (\ref{II9a}) and (\ref{II9b}) in (\ref{II8}) and using (\ref{II8b}) 
the one soliton solution in the explicit form becomes
\begin{eqnarray}
Q(X,T) & = & \frac{L\exp{\eta}}{1 + K\exp{(\eta + \eta ^*)}}\nonumber \\
& = & \frac{L\exp{\{(\mu +i\epsilon)X - (\frac{\mu^3}{6\epsilon} 
+ \frac{\mu \epsilon}{2} + i\frac{\epsilon^2}{3})T\}}}{1 
+ \frac{\mid L\mid ^2}{2\mu ^2}\exp{\{2\mu X - (\frac{\mu^3}{3\epsilon} 
+\mu \epsilon )T\}}}
\label{II10}
\end{eqnarray}
To compare (\ref{II10}) with that of ISM, we choose $L$ as 
\begin{equation}
L = 2\mu \exp{(-i\epsilon X ^{(1)} - \mu X ^{(0)})}
\end{equation}
which reduces (\ref{II10}) to the form
\begin{equation} 
Q(X,T) = \frac{\frac{\mu }{\sqrt 2}
\exp{\{i\epsilon(X -\frac{\epsilon }{3}T-X^{(1)})\}}}
{\cosh{(\mu X -\mu (\frac{\mu^{2}}{6\epsilon} + \frac{\epsilon}{2})T 
-\mu X^{(0)}) +\log{\sqrt {2}})}}
\end{equation}
and this is in agreement with the ISM result of \cite{III}. 

Two Soliton solution of the Sasa Satsuma equation may be obtained, following 
\cite{X}, by choosing G and F of the form 
\begin{subequations}
\begin{equation} 
G = L(\exp{\eta_1} + M\exp{(\eta_1 +\eta_A)})\\
\end{equation}
\begin{equation}
F = 1+k\exp{(\eta_1 + \eta^{*}_1)} + \exp{(\eta_{A})}\\ +kMM^{*}\exp{(\eta_1 
+ \eta^{*}_1 + \eta_{A})}\nonumber\\
\end{equation}
\label{II11}
\end{subequations}
where, $\eta_1$ is complex as in the case of one solition, but $\eta_A$ is 
chosen to be real. Substituting $F$ and $G$ (\ref{II11}) in the bilinear
forms (\ref{II6}), we observe that $\eta_{1}$ soliton satisfies the same 
dispersion relation as (\ref{II8a}), whereas the dispersion relation for 
$\eta_{A}$ becomes $P_{A}^2 = 0$. The degree 2 term in (\ref{II6}a) 
determines the parameter $M$ to be unity. Once again following \cite{X}, we
define the degree of a term by the number of $\eta$'s present in the exponent.
Now the degree 2 and 4 terms in (\ref{II6}b), yield the value of $k$ as in 
(\ref{II8b}). However, because of the additional bilinear form (\ref{II6}c) in
our case, we obtain one more relation, $\epsilon(P_1^* - P_1) = 2i$. 
Note that, in general for complex bosonic systems having complex parameters, the
two soliton solutions may have some relation, which is analogous to the three
soliton conditions \cite{X}. But, the choice of the real parameter $\eta_{A}$, 
makes the three soliton condition trivial in our case. A more general choice of
$F$ and $G$ for two soliton solutions will give such a nontrivial condition.
\vspace {.5 cm}

{\bf III. Riccati Equation and The Conserved Quantities}

To show the Liouville integrability, i.e. the existence of an infinite number of 
conserved quantities related to SSE (\ref{II1}), we first write the associated 
Riccati equation. Since the Lax operators 
in this case are $3\times 3$ matrices, the Riccati equation becomes more 
complicated, though tactable. Let us write the auxiliary field $\Psi $ in the 
component form as
\begin{equation}
\Psi =  \left ( \begin{array}{c}
\chi _{1x} \cr
\chi _{2x} \cr
\chi _{3x}  
\end{array} \right )
\label{III20} 
\end{equation}
Substituting (\ref{III20}) in (\ref{II2}a), we get a set of three coupled 
equations
\begin{subequations}
\begin{eqnarray} 
\chi _{1x} &=& -i\lambda \chi _1 + u \chi _3\\
\chi _{2x} &=& -i\lambda \chi _2 + u ^2 \chi _3\\ 
\chi _{3x} &=& i\lambda \chi _3 - u^*\chi_1 - u\chi _2
\end{eqnarray}
\label{III21}
\end{subequations}
Now expressing (\ref{III21}) in terms of $\Gamma _1 = (\chi _1/\chi_3)$ and 
$\Gamma _2 = (\chi _2/\chi _3)$, and eliminating $\chi _1$, $\chi _2$ and
$\chi _3$, one obtains 
\begin{subequations}
\begin{eqnarray} 
\Gamma _{1x} &=& u - 2i\lambda \Gamma_1 + u ^*\Gamma ^2 _1 + u\Gamma _1 \Gamma _2 \\
\Gamma _{2x} &=& u^* - 2i\lambda \Gamma_2 + u\Gamma ^2 _2 +u^* \Gamma _1 \Gamma _2 
\end{eqnarray}
\label{III22}
\end{subequations}
The first order nonlinear coupled equations (\ref{III22}) for $\Gamma _1$ and $\Gamma _2$ 
are the Riccati 
equations in our case. Notice that neither the integral of $\Gamma _1$ nor
of $\Gamma _2$, plays the role of generating functions for conserved
quantities, but a suitable combination of them does. The infinite number of 
conserved quantities (Hamiltonians), $H_{2n+1}; n=0,1,2,.....$ can be
obtained from (\ref{III22}) by identifying
\begin{subequations}
\begin{equation} 
a (\lambda) = \exp (-i\lambda x)\mid _{x\rightarrow \infty } 
\Psi _3(\infty ,\lambda ) 
 = \exp{\{-\int^{\infty}_{-\infty} (u^* \Gamma _1 + u\Gamma _2 ) dx \}}
\label{III23a}
\end{equation}
where $H_{2n+1}$ are related to $a(\lambda )$ as
\begin{equation}
 \ln a( \lambda) = -2\sum^{\infty} _{n=0} (2i)^{-2n-1}  
H_{2n+1} \lambda ^{-2n-1}
\label{III23b}
\end{equation}
\end{subequations}
We will see that Hamiltonians with odd indices only survive, while the
terms with even indices become trivial. This property is similar to that of the
real KdV or the modified KdV equation. 

We may look for series solutions of (\ref{III22}) by assuming $\Gamma _1$ and 
$\Gamma _2$ in the form
\begin{subequations}
\begin{equation}
\Gamma _1 = \sum ^{\infty} _{n=0} C ^{1} _n \lambda ^{-n}
\end{equation}   
\begin{equation} 
\Gamma _2 = \sum^{\infty} _{n=0}C ^2 _n \lambda ^{-n},
\end{equation}
\label{III24}
\end{subequations}
which yield the following recursion relations from (\ref{III22}a) and 
(\ref{III24}):
\pagebreak

\begin{eqnarray}
C^1_0 = 0 \quad\quad\quad C^1_1 = \frac{u}{2i}\quad\quad\quad\quad\quad\quad\quad\quad\nonumber\\ 
 2i C ^{1} _{n+2} = -(C ^{1}_{n+1}) _x + \sum ^{n+1} _{m= 0} 
(u^{*} C ^{1} _m C ^{1} _{n-m+1} + u C ^{1} _{m} C ^{2} _{n-m+1}). 
\label{III25}
\end{eqnarray}
Similarly (\ref{III22}b) and (\ref{III24}) determine another, though quite 
similar, set of recursion relations given as
\begin{eqnarray}
C ^2_0 = 0 \quad\quad\quad C ^2_1 = \frac{u^*}{2i}\quad\quad\quad\quad\quad\quad\quad\quad\nonumber\\ 
2iC^{2} _{n+2} = -(C ^{2} _{n+1}) _x + \sum^{n+1} _{m=0}
(u C ^{2} _m C ^{2} _{n-m+1} + u^{*}C ^{2} _{m} C ^{1} _{n-m+1})
\label{III26}
\end{eqnarray}
Inserting the expressions of $C^1_n$ and $C^2_n$, thus obtained through the 
recursion 
relations (\ref{III25}),(\ref{III26}) in (\ref{III24}), we get from 
(\ref{III23a},b)
the explicit form of all conserved quantities, $H_{2n+1}$. These expressions 
are the integrals taken over the functions of the fields $u$ and $u^*$ and their 
derivatives. The first few conserved quantities of the infinite set of the SSE 
hierarchies are given by  
\begin{eqnarray}
H_{1} &=& \int u^{*} u dx\\ 
H_{3} &=&  \int (-u ^{*}_x u_x + 2\mid u\mid^4) dx\\ 
H_{5} &=&  \int (u^{*}_{xx} u_{xx} -4\mid u\mid^2 u^*_x u_x 
+ ((\mid u\mid^2)_x)^2 + 8 \mid u\mid^6) dx 
\end{eqnarray}
We have checked explicitly by using equations of motion (\ref{II1}) that $H_1$, $H_3$
and $H_5$ are, indeed, the constants of motion. 
\pagebreak

{\bf IV. Generalised Landau Lifshitz Type Equation as The Gauge Equivalent 
System}

We now show an interesting connection between the SSE in the form (\ref{II1}) 
and the generalized Landau Lifshitz type equation, by exploiting the gauge 
equivalence of the Lax pairs of these two dynamical systems. The procedure is 
similar to that between the NLS and the standard Landau Lifshitz equation 
\cite{VII}. Under a local gauge tranformation, the Jost function, 
$\Psi (x,t,\lambda )$ changes as
\begin{equation}
 {\tilde \Psi }(x,t,\lambda ) = g^{-1}(x,t) \Psi (x,t,\lambda ) 
\label{IV27}
\end{equation}
where $g(x,t) = \Psi (x,t,\lambda )\vert _{\lambda=0}$, may be taken as an 
element of the 
gauge group. Consequently, the Lax equations (\ref{II2}) under this gauge 
transformation (\ref{IV27}) become 
\begin{subequations}
\begin{equation}
{\tilde \Psi }_x = {\tilde {\bf U}}(x,t,\lambda ){\tilde \Psi }
\label{IV28a}
\end{equation}
\begin{equation}
{\tilde \Psi }_t = {\tilde {\bf V}}(x,t,\lambda ){\tilde \Psi }
\label{IV28b}
\end{equation} 
\end{subequations}
where ${\tilde {\bf U}}$ and ${\tilde {\bf V}}$ are the new gauge transformed 
Lax pair, given by
\begin{subequations}
\begin{eqnarray}
{\tilde {\bf U}}(x,t,\lambda ) &=& g^{-1} ({\bf U} - {\bf U}_0) g\\
{\tilde {\bf V}}(x,t,\lambda ) &=& g^{-1} ({\bf V} - {\bf V}_0) g
\end{eqnarray} 
\label{IV29}
\end{subequations}
with 
${\bf U}_0 = {\bf U}\vert _{\lambda = 0} = g_x(x,t)g^{-1}(x,t)$ 
and
${\bf V}_0 = {\bf V}\vert _{\lambda = 0} = g_t(x,t)g^{-1}(x,t)$.

We may identify the spin field of the Landau Lifshitz type equation as
\begin{equation}
S = g(x,t)^{-1} {\bf \Sigma} g(x,t) \quad\quad\quad\quad\quad\quad S^2 = 1.
\label{IV30}
\end{equation}
With this identification, the gauge transformed Lax pair (\ref{IV29}) can be 
expressed in terms of the spin field $S$ (\ref{IV30}) and its derivatives 
only, yielding 
\begin{subequations}
\begin{eqnarray}
{\tilde {\bf U}} &=& -i\lambda S\\
{\tilde {\bf V}} &=& -4i\epsilon \lambda ^3S + 2\epsilon \lambda ^2 SS_x
+ i\epsilon \lambda (S_{xx} + {3\over 2} SS^2_x)
\end{eqnarray}
\label{IV31} 
\end{subequations}
In deriving (\ref{IV31}) one has to use the following important identities\\
$SS_x = 2g^{-1}{\bf A}g$, $SS_x^2 = -4g^{-1}{\bf \Sigma} {\bf A}^2g$ and 
$S_{xx} + SS_x^2 = 
2g^{-1}{\bf \Sigma }{\bf A}_xg$.

The zero curvature condition of (\ref{IV31});
\begin{equation}
{\tilde {\bf U}}_t - {\tilde {\bf V}}_x 
+ [{\tilde {\bf U}} , {\tilde {\bf V}}] = 0
\nonumber
\end{equation}
leads to the generalised Landau Lifshitz type equation
\begin{equation}
S_t + \epsilon S_{xxx} + \frac{3}{2} \epsilon (S_x^3 + SS_{xx}S_x + SS_xS_{xx}) 
= 0
\end{equation}
with $S\in  SU(3)/U(1)$. 

\vspace {.5  cm}
 
{\bf V. Conclusion}

In this paper, we have bilinearised higher order nonlinear Schr\"odinger
equation, vis a vis the SSE following Hirota's method. The 
Hirota's method is an effective and important method to obtain multi soliton 
solutions. We have found explicitly one and two soliton solutions and recovered
the ISM result of one soliton solution from that of Hirota's method after a 
specific choice of the parameter involved. The result related to higher soliton
solutions are more complicated and will be given elsewhere. It is found that 
the SSE falls under a new class of bilinear forms.  

The linear problem of SSE is a nontrivial one in the sense that the Lax operator 
corresponding to this dynamical equation is a $3\times 3$ matrix, which makes the 
related Riccati equation more involved. However, by solving such coupled 
Riccati equations we are able to compute explicitly the infinite number of 
conserved quantities through the recursion relations and to show that terms with 
odd indices only contribute to the conserved charges like KdV and mKdV systems. 
This result establishes explicitly the integrability of the SSE in the Liouville 
sense. Finding out the Poisson bracket structures among the dynamical fields
and consequently revealing the explicit form of the hierarchy of SSE from the 
conserved charges obtained here will be an interesting future problem.

The gauge equivalence of the GLLE with the SSE has been established here. This 
equivalence not only gives a direct relationship between the fields, which 
would help to find the soliton solutions of the GLLE using those of the SSE, but
also yields explicit Lax operators for GLLE from which one would be able to 
extract the related higher conserved quantities in the similar way following the 
present results of SSE.

\end{document}